\documentclass[aps,prd,twocolumn,superscriptaddress,preprintnumbers,showpacs,amsmath,amssymb]{revtex4}
\usepackage{epsf}
\usepackage{latexsym}
\usepackage{amsmath}
\usepackage{amsfonts}
\usepackage{amssymb}
\usepackage{graphicx}
\newcommand{\be}{\begin{eqnarray}}
\newcommand{\ee}{\end{eqnarray}}
\newcommand{\ud}{\mathrm{d}}

\newcommand{\lp}{\ell_{\rm p}}
\newcommand{\mpl}{M_{\rm p}}
\newcommand{\mh}{M_{\rm H}}
\newcommand{\md}{M_{(5)}}
\newcommand{\ld}{\ell_{(5)}}
\newcommand{\mew}{M_{\rm ew}}
\newcommand{\meff}{M_{\rm eff}}
\newcommand{\Meff}{M_{\rm eff}}

\newcommand{\mc}{M_{\rm c}}
\newcommand{\rc}{r_{\rm c}}

\newcommand{\muniv}{M_{\rm univ}}
\newcommand{\rem}{R_{\rm EM}}

\newcommand{\rh}{R_{\rm H}}

\begin{document}
\preprint{Preprint number: DO-TH-10/07}
\title{Affect of brane thickness on microscopic tidal-charged black holes}
\author{Roberto~Casadio}
\email{casadio@bo.infn.it }
\affiliation{Dipartimento di Fisica, Universit\`a di
Bologna and I.N.F.N., via Irnerio 46, 40126 Bologna, Italy}
\author{Benjamin~Harms}
\email{bharms@bama.ua.edu}
\affiliation{Department of Physics and Astronomy, The University
of Alabama, Box 870324, Tuscaloosa, AL 35487-0324, USA}
\author{Octavian~Micu}
\email{octavian.micu@tu-dortmund.de}
\affiliation{Fakult\"at f\"ur Physik, Technische Universit\"at Dortmund,
D-44221 Dortmund, Germany}
\begin{abstract}
We study the phenomenological implications stemming from the dependence
of the tidal charge on the brane thickness $L$ for the evaporation and decay of
microscopic black holes.
In general, the larger $L$, the longer are the black hole life-times
and the greater their maximum mass for those cases in which the black hole
can grow.
In particular, we again find that tidal-charged black holes might live long enough
to escape the detectors and even the gravitational field of the Earth,
thus resulting in large amounts of missing energy.
However, under no circumstances could TeV-scale black holes grow enough
to enter the regime of Bondi accretion.
%
%
\end{abstract}
\maketitle
%
\section{Introduction}
\label{intro}
As is well known by now, the conjectured existence of extra spatial
dimensions~\cite{arkani,RS}
and a sufficiently small fundamental scale of gravity has opened up the possibility
that microscopic black holes can be produced and detected~\cite{bhlhc,CH,BHreview}
at the Large Hadron Collider (LHC).
In a series of papers~\cite{CH,bhEarth1,bhEarth2,bulktidal},
we have analyzed the phenomenology of a particular candidate~\cite{dadhich}
of brane-world black holes~\cite{bwbh} in the Randall-Sundrum (RS) model~\cite{RS}.
Our world is thus a three-brane (with coordinates $x^\mu$, $\mu=0,\ldots,3$)
embedded in a five-dimensional bulk with the metric
\be
\ud s^2 =e^{-|y|/\ell}\,g_{\mu\nu}\,\ud x^{\mu}\,\ud x^{\nu} + \ud y^2
\ ,
\ee
where  $y$ parameterizes the fifth dimension and $\ell$ is a length determined by the
brane tension.
This parameter relates the four-dimensional Planck mass $\mpl$ to the five-dimensional
gravitational mass $\md$ and one can have $\md\simeq 1\,$TeV$/c^2$
(for bounds on $\ell$, see, e.g., Ref.~\cite{harko}) and black holes with mass in the
TeV~range.
The brane must also have a thickness, which we denote by $L$,
below which deviations from the four-dimensional Newton law occur.
Current precision experiments require that $L \lesssim 44\,\mu$m~\cite{Lbounds},
whereas theoretical reasons imply that
$L\gtrsim\ld\simeq \lp\,\mpl/\md\simeq 2\cdot 10^{-19}\,$m ($\lp$ is the
four-dimensional Planck length and, $\ld$ is the five-dimensional Planck length).
In the analysis below, the parameters $\md$ and $L$ are assumed to be
independent of one another, but within the stated ranges.
\par
The tidal-charged metric of Ref.~\cite{dadhich} solves the
Einstein equations projected onto the brane~\cite{shiromizu}
and is given by
\be
\ud s^2 =
- A\,\ud t^2 + A^{-1}\,\ud r^2 + r^2\left(\ud\theta^2 +\sin^2\theta\,\ud\phi^2\right)
\ ,
\label{tidal}
\ee
with
\be
A=1-\frac{2\,\lp\,M}{\mpl\,r}-q\,\frac{\lp^2}{r^2}
\ ,
\ee
where $q>0$ is the tidal charge.
This parameter could  be determined only by solving the Einstein equations in the bulk,
which is a difficult problem whose analytic solution has yet to be obtained
(for a perturbative study, see Ref.~\cite{bulktidal}).
On general grounds, $q$ should depend upon the
Arnowitt-Deser-Misner (ADM) mass $M$ in such a way that $q$ vanishes
when $M\to 0$~\cite{CH,bhEarth1,bhEarth2}.
\par
For specific forms of $q=q(M)$, tidal-charged black holes may have very long
lifetimes~\cite{CH}, which led to the conjecture that they might be able to grow to
catastrophic size within the Earth~\cite{plaga},
contrary to the picture~\cite{giddings} that arises in the ADD scenario~\cite{arkani}.
This possibility was however refuted in Ref.~\cite{GM2}.
We also solved the system of equations which describes the time-evolution of mass
and momentum for various initial conditions and parameters in the acceptable
ranges and found no evidence of catastrophic growth~\cite{bhEarth1,bhEarth2}.
\par
In our previous analysis, we assumed $q=q(M)$, neglecting the possible
dependence of $q$ on the brane thickness $L$.
In the present work, we therefore take a complementary view and assume
that $q=q(M,L)$ with $q=0$ for $L=0$ as well as for $M=0$.
This dependence will be constrained by the experimental bounds mentioned
in the beginning, along with the findings from Ref.~\cite{bulktidal}.
This allows us to restrict the space of parameters to a manageable range,
within which we will study the time-evolution numerically.
We shall then find that tidal-charged black holes produced at
the LHC would very likely evaporate instantaneously and, even for those values
of the parameters which lead to an initial growth, no catastrophic scenario
will arise.
However, life-times are longer for larger $L$ and could be long enough
to allow for black holes to escape from the detectors and result in significant
amounts of missing energy.
\par
We use units with $1=c=\hbar=\mpl\,\lp=\ld\,\md$,
where $\mpl\simeq 2.2\cdot 10^{-8}\,$kg and $\lp\simeq 1.6\cdot 10^{-35}\,$m
are related to the four-dimensional Newton constant $G_{\rm N}=\lp/\mpl$.
Further, $\md\simeq \mew\simeq 1\,$TeV ($\simeq 1.8\cdot 10^{-24}\,$kg),
the electro-weak scale, corresponding to $\ld\simeq 2.0\cdot 10^{-19}\,$m.
\section{Tidal charge and brane thickness}
\label{metrics}
 In some idealized models of the R-S model the brane thickness is taken to be zero.  However this
idealization is not realistic.  In string theory there is a minimum length scale, which implies that there is a
minimum brane thickness.  Furthermore quantum corrections to any classical interactions require that the brane
have a finite thickness.  One simple, although not unique, way of incorporating the condition of finite brane thickness
is to assume that the tidal charge is related to $M$ and $L$ according to
\be
q
\simeq
\left(\frac{L}{\ld}\right)^\gamma\left(\frac{M}{\md}\right)^\beta
\ ,
\label{be_ga}
\ee
where both $\gamma$ and $\beta$ are real and positive.
We further assume that they do not depend on the
mass scale, and can therefore be constrained by using
$\md \simeq 1\,$TeV$/c^2$ and the known bounds on $L$.
For this purpose, we note that the tidal term in the metric
($A=1-A_{\rm t}-A_{\rm N}$),
\be
A_{\rm t}
\simeq
\left(\frac{L}{\ld}\right)^{\gamma}\left(\frac{M}{\md}\right)^\beta
\frac{\lp^2}{r^2}
\ ,
\label{At}
\ee
dominates over the usual Schwarzschild term,
\be
A_{\rm N}\simeq 2\,\frac{M\,\lp}{\mpl\,r}
\ ,
\label{AN}
\ee
for $r\lesssim \rc$, with
\be
\rc\simeq
\lp\,\frac{\mpl}{\md}\,\left(\frac{L}{\ld}\right)^{\gamma}
\left(\frac{M}{\md}\right)^{\beta-1}
\ .
\label{rc}
\ee
Although the thickness of the brane could be much less than the experimental bound on corrections to 
Newton's law of gravitation, we assume the `worst case scenario' in which the thickness of the brane sets the upper
limit on the size of the corrections.  In this way we can push the possibility of catastrophic black hole production at the LHC
to the limit.  Within this scenario consistency requires that $\rc$ be shorter than the length scale
above which corrections to the Newton potential have not yet been detected,
that is
\be
\rc\ll L
\ ,
\label{rcc}
\ee
if the black hole is ``small'', in the sense that
\be
\rh\ll\rc\ll L
\ ,
\label{sbh}
\ee
where
\be
\rh=\lp\left(\frac{M}{\mpl}+\sqrt{\frac{M^2}{\mpl^2}+q}\,\right)
\ ,
\label{rhq}
\ee
is the horizon radius.
For $\rh\ll\rc$, the horizon radius can then be estimated from the
tidal contribution~\eqref{At},
\be
\rh\simeq
\lp\left(\frac{L}{\ld}\right)^{\gamma/2}\left(\frac{M}{\md}\right)^{\beta/2}
\ ,
\label{smallRh}
\ee
otherwise $\rh$ approaches the usual four-dimensional expression
\be
\rh\simeq
2\,\lp\,\frac{M}{\mpl}
\ .
\label{largeRh}
\ee
\par
The effective four-dimensional Euclidean action for such small black
holes is given by~\cite{CH,gergely}
\be
S_{(4)}^{\rm E}
=
\frac{\mpl\,(4\,\pi\,\rh^2)}{16\,\pi\,\lp}
\simeq
\lp\,\mpl\,\left(\frac{M}{\Meff}\right)^{\beta}
\ ,
\label{sS}
\ee
with
\be
\Meff\simeq
{\md}\left(
\frac{\ld}{L}\right)^{\gamma/\beta}
\ .
\ee
According to the area law, a black hole is classical if its mass is much
larger than $\meff$~\cite{mfd,bc2}, which implies that $\meff\lesssim \md$
for the existence of TeV-scale black holes.
Since $\gamma$ and $\beta$ are positive, the above relation holds for
the entire parameter range and stronger constraints could  be imposed only if
$L\lesssim \ld$.
\par
For $\beta\not=1$, one then has that $\rc=L$ corresponds to a critical mass
\be
\mc=
\md\left(\frac{L}{\ld}\right)^{\frac{1-\gamma}{\beta-1}}
\ .
\label{mc}
\ee
Further, for $\beta\not=2$, the condition that $\rc=\rh$ leads to
\be
M\simeq
\mh
\equiv
\md\left[\left(\frac{\md}{\mpl}\right)^{2}
\left(\frac{\ld}{L}\right)^{\gamma}\right]^{\frac{1}{\beta-2}}
\ ,
\label{mh}
\ee
whereas for $\beta=2$ one finds no constraint on $M$, but
$\rh\ll\rc$ implies that
\be
\left(\frac{\mpl}{\md}\right)^{2}\left(\frac{L}{\ld}\right)^{\gamma}\gg 1
\ ,
\ee
which is true for all $\gamma>0$.
\par
For $\beta\neq 1$ and $\beta\neq 2$,  $\mc$ and $\mh$ are
properly defined as above, and the condition~\eqref{rcc} implies
\be
\left(\frac{M}{\md}\right)^{\beta-1}
\ll
\left(\frac{L}{\ld}\right)^{1-\gamma}
\ .
\label{rc_cond}
\ee
We then have two cases:
(i)~for $0<\beta<1$,
\be
M\gtrsim\mc
\ ,
\label{M>Mc}
\ee
while, (ii)~for $\beta>1$,
\be
M\lesssim\mc
\ .
\label{M<Mc}
\ee
Similarly, one can analyze the lower bound in Eq.~\eqref{sbh},
and again find two separate cases:
(a)~for $\beta<2$ we get
\be
M\lesssim\mh
\ ,
\label{betall2}
\ee
and, (b)~for $\beta>2$,
\be
M\gtrsim\mh
\ .
\label{betagg2}
\ee
\subsection{$\beta>1$}
In this range, the tidal term grows with $M$ faster than the Newtonian term, and
Eq.~\eqref{rcc} becomes the upper bound~\eqref{M<Mc} on the maximum black
hole mass, namely
\be
M\lesssim\mc
\ ,
\ee
and the condition $\mc\gg\md$ holds if
\be
0\lesssim\gamma <1
\ .
\label{gamma.0-1}
\ee
We must then analyse the three cases (a), (b) and $\beta=2$ separately.
\subsubsection{$1<\beta<2$}
In this case, a ``small'' black hole must have a mass in the range
\be
\md\ll M\ll \min\{\mh,~\mc\}
\ .
\ee
Requiring that $\mh\gg\md$ implies that
\be
\left(\frac{\mpl}{\md}\right)^{\frac{2}{2-\beta}} \left(\frac{L}{\ld}\right)^{{\frac{\gamma}{2-\beta}}}
\gg 1
\ ,
\ee
which is true for all positive values of $\gamma$, whereas
$\mc\gg \md$ leads to the condition in Eq.~\eqref{gamma.0-1}.
\par
We must however notice that, for $\beta\to 1^+$ and $\gamma$ in the above
range, the critical mass $\mc\to\infty$ and an infinitely massive black hole
would be of the tidal kind, thus ruling out the Schwarzschild geometry.
We therefore require that $\mc\lesssim M_\odot\simeq 10^{54}\,$TeV
(the mass of the sun).
\subsubsection{$\beta=2$}
\label{beta=2}
The black hole mass must be smaller than the critical mass
\be
M\ll\mc
\simeq
\md\left(\frac{L}{\ld}\right)^{1-\gamma}
\ ,
\ee
and $\mc\gg\md$ again leads to Eq.~\eqref{gamma.0-1}.
\subsubsection{$\beta>2$}
The black hole is now ``small'' if
\be
\max\{\mh,~\md\} \ll M\ll \mc
\ .
\ee
The condition that $\mh\ll\mc$ then implies the new constraint
\be
\left(\frac{\md}{\mpl}\right)^{2}
\ll
\left(\frac{L}{\ld}\right)^{\frac{\gamma+\beta-2}{\beta-1}}
\ ,
\ee
which holds for all values of $\gamma$ in the range
given in Eq.~\eqref{gamma.0-1}.
\subsection{$0<\beta<1$}
In this case, the tidal term grows with $M$ more slowly than the Newton
potential, and we obtained
\be
M\gtrsim \mc
\ .
\ee
This also implies that the critical mass $\mc$ must be smaller than $\md$,
which results in the bound~\eqref{gamma.0-1}.
\par
The black hole is then small for
\be
\md\ll M\ll \mh
\ ,
\ee
where $\mh$ is again given in Eq.~\eqref{mh}.
A necessary condition is that $\mh\gg\md$ which holds for $\gamma$
in the range given in Eq.~\eqref{gamma.0-1}.
\subsection{$\beta=1$}
\label{beta1}
Both $A_{\rm t}$ and $A_{\rm N}$ grow linearly with $M$, and~\eqref{rc} reads
\be
\rc\simeq
\lp\,\frac{\mpl}{\md}\,\left(\frac{L}{\ld}\right)^{\gamma}
\ ,
\label{rc1}
\ee
which does not depend on $M$.
Eq.~\eqref{rcc} leads to
\be
\left(\frac{L}{\ld}\right)^{\gamma-1}\ll 1
\ .
\label{beta.1}
\ee
which again constrains $\gamma$ to the range given in Eq.~\eqref{gamma.0-1}.
\section{Time-evolution}
\label{evol}
The time evolution of the black hole mass is obtained by summing
the evaporation and accretion rates~\cite{giddings,bhEarth2},
\be
\frac{\ud M}{\ud t} =
\left.\frac{\ud M}{\ud t}\right|_{\rm evap}
+\left. \frac{\ud M}{\ud t}\right|_{\rm acc}
\ .
\label{dMdt}
\ee
Evaporation occurs via the Hawking effect~\cite{hawking} and is described in
the microcanonical picture~\cite{mfd,bhEarth1}.
There are two mechanisms by which microscopic black holes accrete mass:
one due to the collisions with the atomic and sub-atomic particles encountered
as they sweep through matter, and one due to the gravitational force the black
hole exerts on surrounding matter once it comes to rest.
The latter is known as Bondi accretion and is appreciable only when the
horizon radius is greater than atomic size~\cite{bondi}.
Eq.~\eqref{dMdt} and the equation for the time-evolution of the momentum,
\be
\frac{\ud p}{\ud t}
=
\frac{p}{M}\left.\frac{\ud M}{\ud t}\right|_{\rm evap}
\ .
\label{dpdt}
\ee
form a system which can be solved numerically to obtain $M(t)$ and $p(t)$.
\par
In the following we shall evolve a black hole produced with
a typical initial mass $M(0)\simeq 10\,$TeV$/c^2\simeq 10^{-23}\,$kg
and momentum $p(0)$ varied in the range from $1\,$MeV$/c$
to $5\,$TeV$/c$~\footnote{These
values correspond to a black hole energy in the laboratory of about $11\,$TeV
and are chosen considering the LHC total collision energy of $14\,$TeV and
the fact that a black hole cannot be the only product of a collision.}.
We will analyze values for the parameter $\beta>0$ in each of the different ranges
considered in Section~\ref{metrics} and
\be
0\lesssim\gamma< 1
\ ,
\ee
The brane thickness $L$ will be varied in the range
$10^{-13}\,\mu$m$\,\lesssim L\lesssim 44\,\mu$m, with
the corresponding critical mass $\mc$ given in Eq.~\eqref{mc}.
Our results are given in Tables~\ref{p_0_large}-\ref{m_of_beta}, in which
$\muniv$ is the black hole mass after a time of the same order
of magnitude as the present age of our Universe ($10^{18}\,$sec), $\rh$ and
$\rem$ the corresponding values of the horizon and capture radius,
$M_{\rm E}$ the mass reached after traveling the Earth's diameter,
$R_{\rm E}$ the corresponding capture radius, $t_{\rm E}$ the time
to travel the Earth's diameter and $v_{\rm E}$ the velocity at that point.
\begin{table}
\centering
\begin{tabular}{|c|c|c|c|c|c|}
\hline
$p(0)$ & $M(0)$  &$M_{\rm E}$ & $R_{\rm E}$ & $t_{\rm E}$ & $v_{\rm E}$
\\
(TeV$/c$) & (TeV$/c^2$) & (kg) & (m) & (sec) &
(km$/$sec)
\\
\hline
$5.0$&$10$& $1.8\cdot 10^{-23}$ &$1.3\cdot 10^{-18}$ &$0.1$ &$1.3\cdot 10^{5}$
\\
\hline
$1.0\cdot 10^{-1}$&$11$& $2.0\cdot 10^{-23}$ &$1.3\cdot 10^{-18}$ &$5.2$ &$2.6\cdot 10^{3}$
\\
\hline
$1.0\cdot 10^{-2}$&$11$& $2.0\cdot 10^{-23}$ &$1.3\cdot 10^{-18}$ &$52$ &$2.6\cdot 10^{2}$
\\
\hline
$1.0\cdot 10^{-3}$&$11$& $2.0\cdot 10^{-23}$ &$1.3\cdot 10^{-18}$ &$4.9\cdot 10^{2}$ &$26$
\\
\hline
$1.0\cdot 10^{-4}$&$11$& $2.0\cdot 10^{-23}$ &$1.3\cdot 10^{-18}$ &$5.2\cdot 10^{3}$ &$2.6$
\\
\hline
\end{tabular}
\caption{Time-evolution of black hole mass for {\em large\/}
initial momentum $p(0)$.
$L=10^{-4}\,\mu$m, $\beta=1.1$, $\gamma=0.1$ which result in
$M_c=4\cdot 10^{54}\,$TeV$/c^2$.
\label{p_0_large}}
\end{table}
\begin{table}
\centering
\begin{tabular}{|c|c|c|c|c|}
\hline
$p(0)$ & $M(0)$  & $\muniv$ & $\rem$ & $\rh$
\\
(MeV$/c$) & (TeV$/c^2$) & (kg) & (m) & (m)
\\
\hline
$100$&$11.0$& $8.4\cdot 10^{-15}$ & $2.0\cdot 10^{-16}$&$9.2\cdot 10^{-30}$
\\
\hline
$10$&$11.0$& $1.8\cdot 10^{-15}$ & $1.3\cdot 10^{-16}$&$3.9\cdot 10^{-30}$
\\
\hline
$1.0$&$11.0$& $3.9\cdot 10^{-16}$ & $9.4\cdot 10^{-17}$&$1.7\cdot 10^{-30}$
\\
\hline
\end{tabular}
\caption{Time evolution of black hole mass for {\em small\/} initial momentum
$p(0)$.
$L=10^{-10}$m, $\beta=1.1$, $\gamma=0.1$ which result in
$M_c=4\cdot 10^{54}\,$TeV$/c^2$.
\label{p_0_small}}
\end{table}
\begin{table}
\centering
\begin{tabular}{|c|c|c|c|c|c|}
\hline
$L$ & $\mc$  & $\muniv$ & $\rem$ & $\rh$
\\
(m) & (TeV$/c^2$) & (kg) & (m) & (m)
\\
\hline
$10^{-10}$& $4\cdot 10^{54}$
&$8.4\cdot 10^{-15}$ & $2.0\cdot 10^{-16}$&$9.2\cdot 10^{-30}$
\\
\hline
$10^{-11}$& $4\cdot 10^{45}$
&$3.0\cdot 10^{-15}$ & $9.1\cdot 10^{-17}$&$5.4\cdot 10^{-30}$
\\
\hline
$10^{-12}$& $4\cdot 10^{36}$
&$1.8\cdot 10^{-15}$ & $4.2\cdot 10^{-17}$&$3.1\cdot 10^{-30}$
\\
\hline
$10^{-13}$& $4\cdot 10^{27}$
&$8.4\cdot 10^{-16}$ & $1.9\cdot 10^{-17}$&$1.8\cdot 10^{-30}$
\\
\hline
$10^{-14}$& $4\cdot 10^{18}$
&$3.9\cdot 10^{-16}$ & $9.1\cdot 10^{-18}$&$1.0\cdot 10^{-30}$
\\
\hline
$10^{-15}$& $4\cdot 10^{9}$ &
N/A & N/A&N/A
\\
\hline
\end{tabular}
\caption{Time evolution of black hole mass as function of brane thickness
$L$ for $\beta=1.1$, $\gamma=0.1$ and
initial conditions $M(0)=11\,$TeV$/c^2$ and $p(0)=100\,$MeV$/c$.
N/A means that the black hole mass does not grow.
\label{L_dependence}}
\end{table}
\begin{table}
\centering
\begin{tabular}{|c|c|c|c|c|c|}
\hline
$\gamma$ & $\mc$  & $\muniv$ & $\rem$ & $\rh$
\\
 & (TeV$/c^2$) & (kg) & (m) & (m)
\\
\hline
$0.1$& $4\cdot 10^{54}$
&$8.4\cdot 10^{-15}$ & $2.0\cdot 10^{-16}$&$9.2\cdot 10^{-30}$
\\
\hline
$0.3$& $2\cdot 10^{37}$
&$8.4\cdot 10^{-15}$ & $2.0\cdot 10^{-16}$&$9.2\cdot 10^{-30}$
\\
\hline
$0.5$& $4\cdot 10^{54}$
&$8.4\cdot 10^{-15}$ & $2.0\cdot 10^{-16}$&$9.2\cdot 10^{-30}$
\\
\hline
$0.7$& $4\cdot 10^{2}$
&N/A&N/A&N/A
\\
\hline
\end{tabular}
\caption{Time evolution of black hole mass as a function of $\gamma$ for
$\beta=1.1$, $L=10^{-4}\,\mu$m and
initial conditions $M(0)=11\,$TeV$/c^2$ and $p(0)=100\,$MeV$/c$.
N/A means that the black hole mass does not grow.
\label{m_of_gamma}}
\end{table}
\begin{table}
\centering
\begin{tabular}{|c|c|c|c|c|c|}
\hline
$\beta$ & $\mc$  & $\muniv$ & $\rem$ & $\rh$
\\
 & (TeV$/c^2$) & (kg) & (m) & (m)
\\
\hline
$1.1$& $4\cdot 10^{54}$
&$8.4\cdot 10^{-15}$ & $2.0\cdot 10^{-16}$&$9.2\cdot 10^{-30}$
\\
\hline
$1.2$& $4\cdot 10^{2}$
&N/A&N/A&N/A
\\
\hline
\end{tabular}
\caption{Time evolution of black hole mass as function of $\beta$ for
 $\gamma=0.1$, $L=10^{-4}\,\mu$m and
initial conditions $M(0)=11\,$TeV$/c^2$ and $p(0)=100\,$MeV$/c$.
N/A means that the  black hole mass does not grow.
\label{m_of_beta}}
\end{table}
\subsection{Rapidly decaying solutions}
The first important result is that, for $0<\beta<1$ and $1.3\lesssim\beta$
(this lower bound slightly varies depending on the values of
$\gamma$ and $L$), the black hole decays almost instantly.
In fact, the decay time is less than $10^{-10}\,$sec and
the black hole would not exit the detector inside which it was produced.
\subsection{Growing solutions}
\begin{figure}[t]
\centering
\epsfxsize=3.0in
\epsfbox{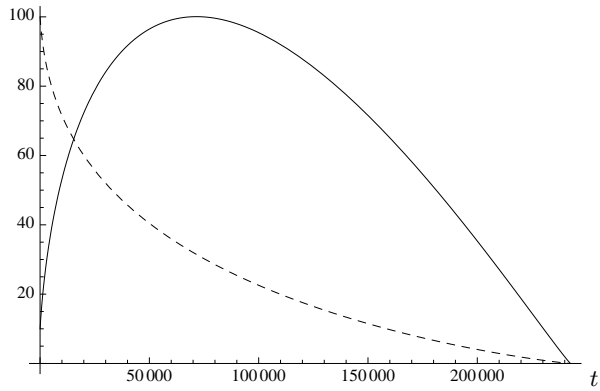}
$t$
\caption{Mass (TeV$/c^2$; solid line) and momentum (MeV$/c$; dashed line)
for $L=5\,\mu$m, $\beta=1.3$, $\gamma=0.01$, $M(0)=10\,$TeV$/c^2$
and $p(0)=100\,$MeV$/c$.
(Time in seconds.)}
\label{slow}
\end{figure}

For $1<\beta\lesssim 1.3$, the mass of the black hole can grow and
a typical example is displayed in Fig.~\ref{slow}.
\par
\begin{figure}
\centering
\epsfxsize=3.0in
\epsfbox{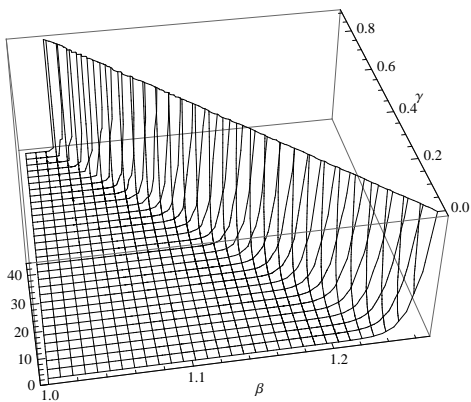}
\caption{Plot of $L=L(\beta,\gamma)$ (in $\mu$m) for $\mc=M_\odot$.
Acceptable values lie below displayed surface.}
\label{fig_mc}
\end{figure}
In Table~\ref{p_0_large}, we show the evolution of the masses of the black holes
as a function of the initial momentum $p_0$ for $\gamma=0.1$, $\beta=1.1$ and
$L=10^{-4}\,\mu$m.
The critical mass in this case is constant and approximately equal to its maximum
allowed value $4\cdot 10^{54}\,$TeV$/c^2$.
The black hole velocity $v_E$ after traveling a distance equal to the Earth's diameter
is larger than the Earth's escape velocity for all values of the initial momentum
higher than $1\,$GeV$/c$.
This means that, regardless of their initial trajectories, the black holes would escape
from the Earth, at which point accretion would turn off, and they would only evaporate.
\par
In Table~\ref{p_0_small} we turn our attention to black holes which are produced
with an initial momentum small enough to get trapped within the Earth.
We evolve the black hole mass for $10^{18}\,$sec, which is comparable
to the present age of our Universe.
We need to emphasize that black holes which are trapped inside the Earth
will decay in times much smaller than this, since at the top of the trajectory
their velocity goes to zero in the radial direction.
At this point accretion stops and the black hole evaporates rapidly.
The data we present simulates a scenario in which the whole Universe
is filled with matter of density equal to the Earth's.
We notice that the final mass increases with the value of the initial momentum.
The highest attainable mass in this time frame is of the order of $10^{-14}\,$kg,
for which both the electromagnetic radius and horizon radius are orders of
magnitude smaller than atomic size.
Bondi accretion therefore does not occur.
\par
Table~\ref{L_dependence} shows the dependence of the mass $M$ on
the brane thickness up to $L=10^{-4}\,\mu$m, since for $L$ larger, the critical mass
would be greater than the mass of the sun.
The data shows that $\mc$, $\muniv$, $\rem$, and $\rh$ all increase
with $L$ and, for $L\lesssim 10^{-9}\,\mu$m the black holes evaporate instantly.
The maximum values of $\rem$ and $\rh$ are again many orders of magnitude
smaller than atomic size and the black holes remain far from Bondi accretion.
We note here that the the range $10^{-4}\,\mu$m$\,\ll L$ is not excluded.
The parameter $\gamma$, for instance, can be slightly increased in order to keep
$\mc$ within limits for all values of $L$ smaller than $44\,\mu$m,
as one can see from Fig.~\ref{fig_mc}.
Nevertheless, $\mc$, $\rem$, and $\rh$ will not vary more than two orders of
magnitude from the values presented in Table~\ref{L_dependence}.
\par
The study of the dependence on $\gamma$ does not show strong variations in
the final values at $10^{18}\,$sec.
Table~\ref{m_of_gamma}  shows that increasing $\gamma$ to $0.7$ still leads to
instantaneous decay.
The disappearance of a peak in $M(t)$ is more sensitive to $\beta$, as can be seen
from Table~\ref{m_of_beta}.
\section{Conclusions}
\label{conc}
In this investigation we have continued our study of the microscopic black holes which could be
produced at the LHC, based on the model presented in Refs.~\cite{CH,bhEarth1}
and the description of brane-world black holes given in Ref.~\cite{dadhich}.
In particular, we have extended the treatment of Ref.~\cite{bhEarth2} by allowing
the tidal charge to depend on the brane thickness.
Conditions for the black holes to be tidal and phenomenologically acceptable
were then used to determine the range of the parameters $\gamma$ and $\beta$
in Eq.~\eqref{be_ga}.
Subsequently, the time evolution of the black hole mass and momentum
were obtained numerically.
\par
We find that tidal black holes would evaporate (almost) instantly,
except for $1<\beta\lesssim 1.3$.
Inside this range, black holes cannot grow to catastrophic size,
but might live long enough to escape the detectors and result in significant amounts of missing energy.
\end{document}